\documentclass{article}
\usepackage[utf8]{inputenc}
\usepackage{makeidx}  % allows for indexgeneration
\usepackage{graphicx}
\usepackage{balance}
\usepackage{listings}
\usepackage{multirow}
\usepackage{threeparttable}
\usepackage{booktabs}
\usepackage{epstopdf}
\usepackage{amsmath,amsfonts,amssymb,mathrsfs,amsthm}
\usepackage{float}                              %per imporre QUI con [H]
\usepackage{multirow}
\usepackage{caption}
\usepackage{subfig}
\usepackage{footnote}
\usepackage{color}
\usepackage{acronym}
\usepackage{pgf-pie}
\usepackage{xcolor}

\acrodef{PV}{photovoltaic}
\acrodef{HEMS}{home energy management system}
\acrodef{NILM}{non-intrusive load monitoring}
\acrodef{HMM}{hidden Markov model}
\acrodef{FHMM}{fractional hidden Markov model}
\acrodef{PF}{particle filtering}
\acrodef{ILM}{intrusive load monitoring}
\acrodef{FSM}{finite state machine}
\acrodef{RMSE}{root mean square error}

\usepackage{tikz}
\usepackage{pgfplots}
\pgfplotsset{width=7cm,compat=newest}

\definecolor{orange}{rgb}{1,0.5,0}

\makeatletter
\def\blfootnote{
 \xdef\@thefnmark{}\@footnotetext
 }
\makeatother

\begin{document}

\lstset{ %
 % language=Prolog,                % the language of the code
  basicstyle=\ttfamily\scriptsize,         % the size of the fonts that are used for the code
  numbers=left,                   % where to put the line-numbers
  stepnumber=0,                   % the step between two line-numbers. If it's 1, each line
                                  % will be numbered
}

\title{Autonomous Load Disaggregation Approach based on Active Power Measurements}

 \author{
Dominik~Egarter, and~Wilfried~Elmenreich\\
Institute of Networked and Embedded Systems\\
 Alpen-Adria-Universit\"at Klagenfurt, Austria\\
 \{\emph{name.surname}\}@aau.at
 } 

\maketitle              % typeset the title of the contribution

\begin{abstract}
With the help of smart metering valuable information of the appliance usage can be retrieved.
In detail, \ac{NILM}, also called load disaggregation, tries to identify appliances in the power draw of an household.
In this paper an unsupervised load disaggregation approach is proposed that works without a priori knowledge about appliances.
The proposed algorithm works autonomously in real time.
The number of used appliances and the corresponding appliance models are learned in operation and are progressively updated.
The proposed algorithm is considering each useful and suitable detected power state.
The algorithm tries to detect power states corresponding to on/off appliances as well as to multi-state appliances based on active power measurements in $1s$ resolution.
We evaluated the novel introduced load disaggregation approach on real world data by testing the possibility to disaggregate energy demand on appliance level.\newline
\end{abstract}
\section{Introduction}

%General Issue

The power draw of a household can potentially reveal a lot of information regarding the used devices, their individual power draw and behavioral patterns of the user(s). While this can constitute a severe privacy problem~\cite{prokop:14}, this information can be also used locally to analyze the usage and power consumption of devices in order to provide information for energy counseling, energy management applications, and increasing energy awareness to the user by providing detailed device-level feedback~\cite{monacchi2013Nov}. While we expect a raising number of smart appliances~\cite{elmenreich:wises12} in the future, a considerable number of household appliances will be legacy devices which are not able to directly report their operational data regarding time and consumption. Using a high number of dedicated meters to monitor these devices will be neither cost nor energy-effective. Non-intrusive load monitoring overcomes this problem by applying a single meter approach to acquire a time series of power measurements which are then processed in order to infer about the used appliances. However, many of such load disaggregation algorithms require previous knowledge about the devices employed in the system.

In this paper we present an unsupervised load disaggregation approach that is able to identify device operations based on the characteristic power changes when devices are switched on/off or switch to a different power state. Given, that power states of devices are distinguishable, the proposed algorithm does not need {\em a priori} information about the system and autonomously adapts to new or removed devices. The algorithm can be used online and is suitable for operation on low-cost embedded system hardware, for example as part of an energy management system.

The presented approach constitutes an important step towards an automatic disaggregation of electrical loads. The approach is especially suitable for household appliances, since these environments feature typically different power draws out of device pool that is also subject to change over a larger timescale by acquisition of new devices. By presenting a working approach for automatizing the detection of devices without supervision, i.e., without the need for querying the user every time the device pool has changed, this paper lays the ground for a broad application of load disaggregation.

The following section gives an overview on related work on load disaggregation, depicts the problem statement and describes our approach. The particular steps of the algorithm are explained in Section~\ref{sec:algorithm} in detail. The approach has been evaluated in a case study based on available household consumption datasets. Limitations and future work are discussed in Section~\ref{sec:limitations} before the paper is concluded in Section~\ref{sec:conclusion}.

%Outline

\section{Background and Approach}

\subsection{Related Work}\label{subsec:relatedwork}
The first approach of \ac{NILM} was introduced by G.Hart \cite{Hart1992}.
He used active and reactive power readings to establish appliance models based on \ac{FSM} which he  used to infer an appliance to be on or off.
Current approaches solving the load disaggregation problem can be distinguished between supervised and unsupervised approaches.
A good overview on supervised approaches are described in \cite{Zeifman2011} and \cite{Zoha2012}.
In the following we are focusing on unsupervised load disaggregation approaches.
Unsupervised classification approaches do not need any {\em a priori} information of the system.
In particular, no labelled data is needed to learn models and to perform classification.
Recent approaches are based on \acp{HMM} \cite{Zia2011}, on \acp{FHMM} \cite{Zoha2013}, on different variations of \acp{HMM} \cite{Zico2012,Kim2011} and on temporal motif mining \cite{Shao2013}.
For these approaches the distinction between appliances is unsupervised whereas the labelling which model corresponds to which appliance is done not automatic.
Approaches performing automatic labelling is conducted based on Bayesian inference \cite{Johnson2013} and on an semi-supervised classification \cite{Parson2014}.
The most related approach to the presented work is presented in \cite{pattem2012}, which provides an unsupervised magnitude-based disaggregation approach based on \ac{HMM}s.
Our work is different to this work in several aspects as the algorithm performs online, is autonomous and self-learning, considers all possible appliance states and needs no learning of appliance transitions which often is subject to erroneous observations.
Furthermore, we tested our approach on aggregated data from known appliances which makes the comparison on appliance level power states with the detected power states and inferred power states possible.
More information about the evaluation process is presented in \ref{sec:evaluation}.

\subsection{Problem Statement}\label{subsec:problemStatements}
The problem to disaggregate appliance readings from the aggregated power draw is composed by overlapping appliance power draws, where each appliance has a power draw $p_i(t)$ and the aggregated power $P(t)$ can be formulated as the sum of each appliance's power consumption:
\begin{equation}
P(t) = \sum^N_{i=1} p_i(t)
\end{equation}
The variable $N$ represents the number of used appliances.
Current research approaches as presented in Section \ref{subsec:relatedwork} are focusing on unsupervised load disaggregation approaches.
The amount of {\em a priori} information should be minimized without a reduction of the information gain produced by load disaggregation.
Without any {\em a priori} information several problems arise for an load disaggregator and have to be considered:
\begin{itemize}
\item The number of used appliances has to be identified.
Current clustering approaches need to know the number of used appliances
\item The appliance model has to be learned without any {\em a priori} information and in operation.
New appliances should be added to the load disaggregation approach and rarely used appliances should be deleted from the set of appliances used by the load disaggregator.
\item Suitable appliance features should be extracted from noisy and low frequency active power readings. The detected features are used to generated meaningful appliance models. The load disaggregator should be working based on power magnitudes.
\end{itemize}
Furthermore, the algorithm should work online.
This means on the one hand supporting model learning in operation and on the other hand being capable of making appliance state estimations based on the learned models in real time.
Therefore, the computational effort of the approach should be bound (for a reasonable set of appliance models) and match the performance of state-of-the-art embedded hardware.

\subsection{Basic Approach}
The proposed load disaggregation approach considers the presented problems of Section \ref{subsec:problemStatements} and is performing autonomously and in an unsupervised way.
No {\em a priori} information as the number of appliances or appliances informations is needed.
The load disaggregation approach is usable with a minimal amount of power reading informations.
Our proposed approach can be divided into four processing stages which are visualized in Figure \ref{fig:overview}.
\begin{itemize}
\item \textbf{State Detection}: Aims to detect significant power edges which can be assigned to appliance switching events. Data preprocessing as signal smoothing and de-noising takes place at this processing stage.
\item \textbf{State Clustering}: Power edges are formed to state clusters to identify the most important states or switching events. These states are used to create appliance models used by the load disaggregator.
\item \textbf{Classification}: With the appliance models generated, appliance states should be estimated by an online load disaggregation approach using low frequency active power readings.
\item \textbf{Appliance Database Update}: To add, to maintain and to delete appliance models in an autonomous way, this stages is responsible to find new power states, to improve the power states of existing appliance models and to delete appliance models which appeared only once or very rarely.
\end{itemize}

\section{Autonomous Load Disaggregation} \label{sec:algorithm}

\begin{figure*}
\centering
\includegraphics[width=1\columnwidth]{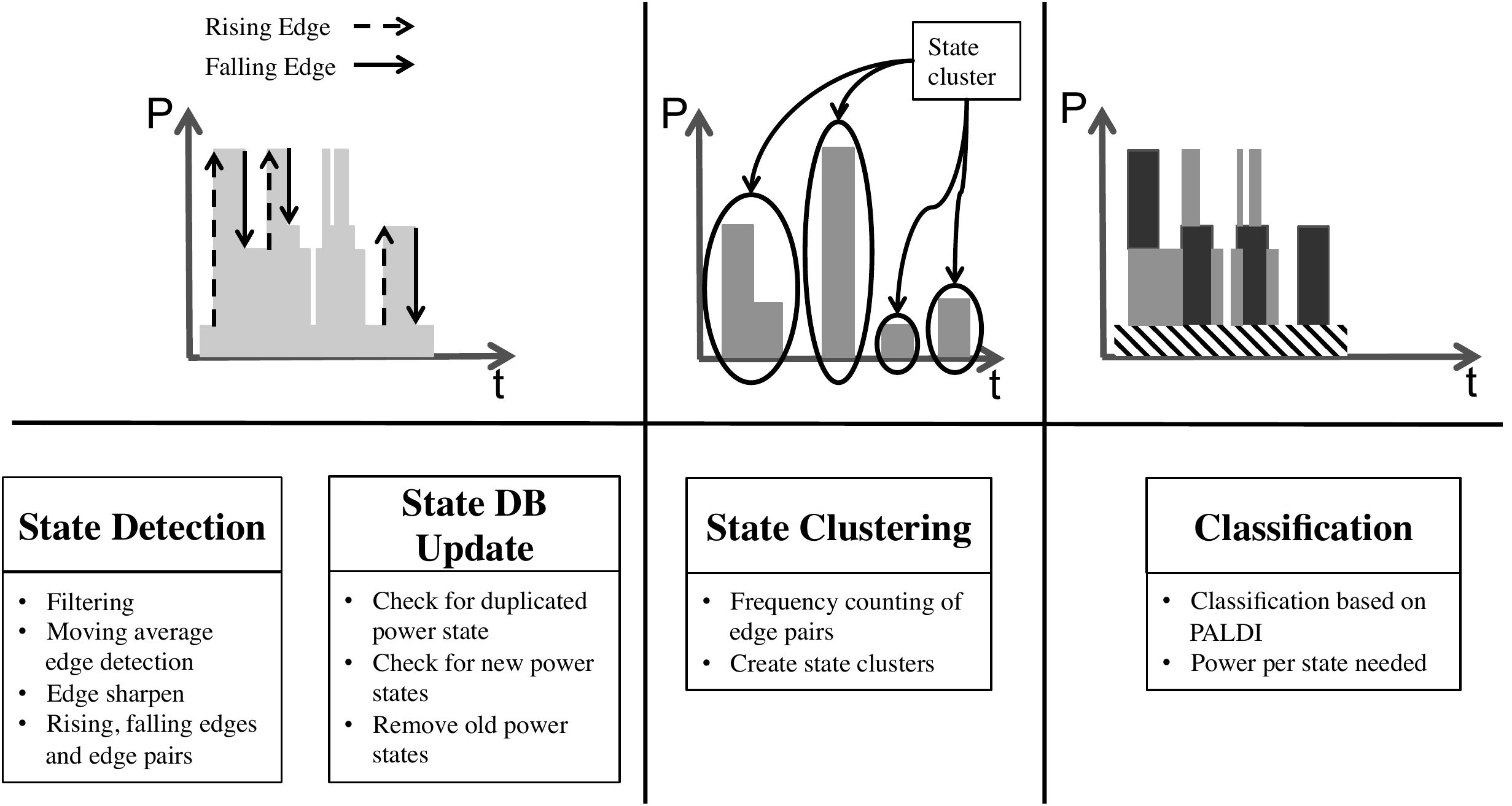}
\caption{General computation sequence of the unsupervised disaggregation approach. At the first stage the power draw is preprocessing by filtering and smoothing the signal to perform edge detection. Detected rising and falling edges are paired together trying to find searched appliance power states. Power states are used to model appliances, which are finally used to disaggregated the appliance power draws from the total demand.}
\label{fig:overview}
\end{figure*}

The presented load disaggregator is autonomous and unsupervised.
No {\em a priori} knowledge about the number and power value of appliances in the system is needed. The used appliance models are created and updated in operation and are used by the load disaggregator to make estimates which detected appliance was used at which point in time.
A system overview of all processing stages is presented in Figure \ref{fig:overview} including \textit{state detection}, \textit{State DB update}, \textit{state clustering}, and \textit{clustering}.
In the following each processing stage is described in detail.

\subsection{Feature Detection}
One of the major tasks of the proposed load disaggregation approach is to detect and to identify useful appliance features.
According to our assumption we focus on smart metering readings of active power ratings with a measurements resolution of $1Hz$.
With the aggregated power readings we aim to extract appliance features based on appliance switching events.
In detail, we concentrate on switching ON and switching OFF events where all power states of an appliance are taken under consideration.
The task is to produce abrupt edges with a significant change without losing important appliance related information in which power transients can last several seconds in real.
Due to the fact the measurement readings are affected by noise, the reading have to be preprocessed to get sufficient and satisfying data.
Thus, we de-noise the power readings by median filtering with an appropriate window size of for example 30 readings at a measurement frequency of 1 Hz.
The window size has to chosen carefully since a window chosen too wide could lead to information loss by wiping out important edges.
Edge detection based on moving average and thresholding is applied on the filtered power readings to detect rising as well as falling edges.
All rising and falling edges are checked for matching pairs to create a pool of possible appliance power states.
This processing stage is performed on sliding time window of predefined size.
We empirically identified one day as a suitable time window balancing a fair amount of switching events needed by processing stages.
\subsection{State Clustering and Appliance Creation}
The pool of occurred power edges is the basis for the following analysing process which aims to create appliance models used be the final classifier.
To create appliance models, the first task is create a histogram of all edge pairs detected by the feature detectors.
The created histogram counts the occurred power edges from $0$ to $3000W$ each $5W$.
This task is followed by a segmentation stage.
Segmentation is used as a form of clustering to combine similar occurred edges to one edge pair representing a possible appliance power state.
The set of possible power states is used to create appliance models including their nominal power consumption in operation.
The appliance models are saved in a simple database which is updated in use.
In detail, new appliances according to newly occurred power states are included and appliance models and power states which occur rarely are removed from the appliance model database.
We model each appliance as an \ac{HMM} described by an initial state, by its transition matrix, and by its observations matrix.
The detected power states are assigned to the observation matrix of a \ac{HMM} representing the appliance power demand in operation.
The off state ($0W$) is assigned to each appliance \ac{HMM} as second observation and as the initial state .
The total power demand is modelled by an \ac{FHMM} where the set of appliance \ac{HMM}s aggregate their power observations over time.

\subsection{Online Classification}
Our appliance classifier is based on Baysian inference.
We use the online load disaggregator presented in \cite{Egarter2014}.
The approach is based on \ac{PF} aiming to approximate the posterior density of the \ac{FHMM}
The approach disaggregates each appliance power demand and appliance state from the household demand, according to the current observed consumption and the given appliance models.
The \ac{PF} output estimates the household consumption and inputs this information and the appliance models to a simple decision maker based on thresholding.
The use of a \ac{PF} as load disaggregator is beneficial for three reasons.
First, \ac{PF} can handle non-linear problems presented by non-linear behaving loads such as a driller or a dimmer.
Second, it can handle non-Gaussian noise influences resulting from uncertainty in power trends and consumption data.
Third, \ac{PF} and its performance can be adjusted by the number of used particles.
The more particles the \ac{PF} considers, the better the estimated posterior density.
The number of particles is limited by the computational effort of the approximation process.
We empirically identified 1000 particles as an appropriate number balancing the trade-off between the context of computational effort and detection performance.
Exact knowledge of the transition matrix values is not necessary since the \ac{PF} is independently estimating the appliance states by an appropriate number of used particles.
In case of a two-state appliance represented by a two-state transition matrix, a clear trend which state is more probable than the other should be visible.
This simplifies the appliance modelling stage and makes the approach of \cite{Egarter2014} usable for the appliance models employed in this work.
The disaggregation process is performed on each measurement sample (each second) and considers only the current power sample by the estimation process.

\subsection{Appliance Database Update}
In each time window power edges and appliances are generated.
It is obvious that one time window is not representative for an set of appliances.
Appliances are used in different times and days as wells as repetitions.
Thus, the process to generate appliances has to be performed on each time window which rises the need to update saved appliance models.
We implemented an updating process which is checking for new appliances and power states, for appliances or power states which are  similar to existing appliances and for appliances which are only rarely used.
We are tracking appliance usage meta data including power state, appearance per day, power estimates and operational time for each day.
The parameters are used to update the saved appliance model database.
Moreover, we use a threshold of $50W$ to distinguish between two appliances.
Thus, appliances with a power draw indistinguishable from each other with respect to the measurement accuracy of the system are modeled both as the same virtual appliance with the respective power draw.

\section{Evaluation} \label{sec:evaluation}

\subsection{Implementation \& Evaluation Settings}
We implemented the presented unsupervised approach in Matlab.
As input of the approach we used an aggregated consumption dataset based on measurements from real households.
No further input is forwarded the approach.
The presented tests and evaluations are simulation based and were run on a MacBook Pro, 2,8GHz with 8GB RAM.
To test the approach according to its detection performance we used the disaggregation error on appliance level describing the \ac{RMSE} of the estimated and ground truth power data.
Due to the fact that the unsupervised approach is based on unlabelled data, the labelling of the power states and appliances was made empirically.
Each detected appliance is mapped to a known power state of the ground truth or to the "\textit{unknown}" appliance state container.
The "\textit{unknown}" appliance state container presents appliance states which were not previously identified by a human observer but were detected by the algorithm.
Moreover, similar appliance of the ground truth data are combined as one virtual appliance since the algorithm has no possibility to distinguish between appliances with the same power demand.

\subsection{Dataset}
There exists several publicly available datasets which can be employed to the test our proposed approach such as the REDD dataset \cite{kolter-kdd-2011}, the GREEND dataset \cite{Andrea2014}, the ECO-dataset \cite{Beckel2014}, the AMPds dataset \cite{makonin2013ampds}, and the Smart* dataset \cite{barker2012smart}. We chose the REDD dataset as reference because of the recording parameters and due to its wide application as a standard test set.
This dataset offers active power readings on appliance level at approximately 1Hz resolution for 6 different houses.
We took one house $1$ of the dataset for evaluation using six appliances (oven, fridge, dishwasher, kitchen outlet, microwave, washing dryer) to generate the aggregated power load.
The appliances were chosen based on to their contribution to the household power demand \cite{Carlson2013132}.
As time period we took 30 consecutive days.

\subsection{Case Study}

%\subsubsection{Preprocessed Data}
%\begin{figure}[htpb]
%\begin{minipage}[hbt]{0,48\textwidth}
%		\newlength\figureheight
%		\newlength\figurewidth
%		\setlength\figureheight{4cm}
%		\setlength\figurewidth{7cm}
%		\input{preprocessed.tikz}
%		\caption{Number of power events detected and assignable to }
%		\label{fig:detectedStates}
%\end{minipage}
%\end{figure}

\subsubsection{Number of Detected Power States per Day}
\begin{figure}
\centering
\includegraphics[width=1\columnwidth]{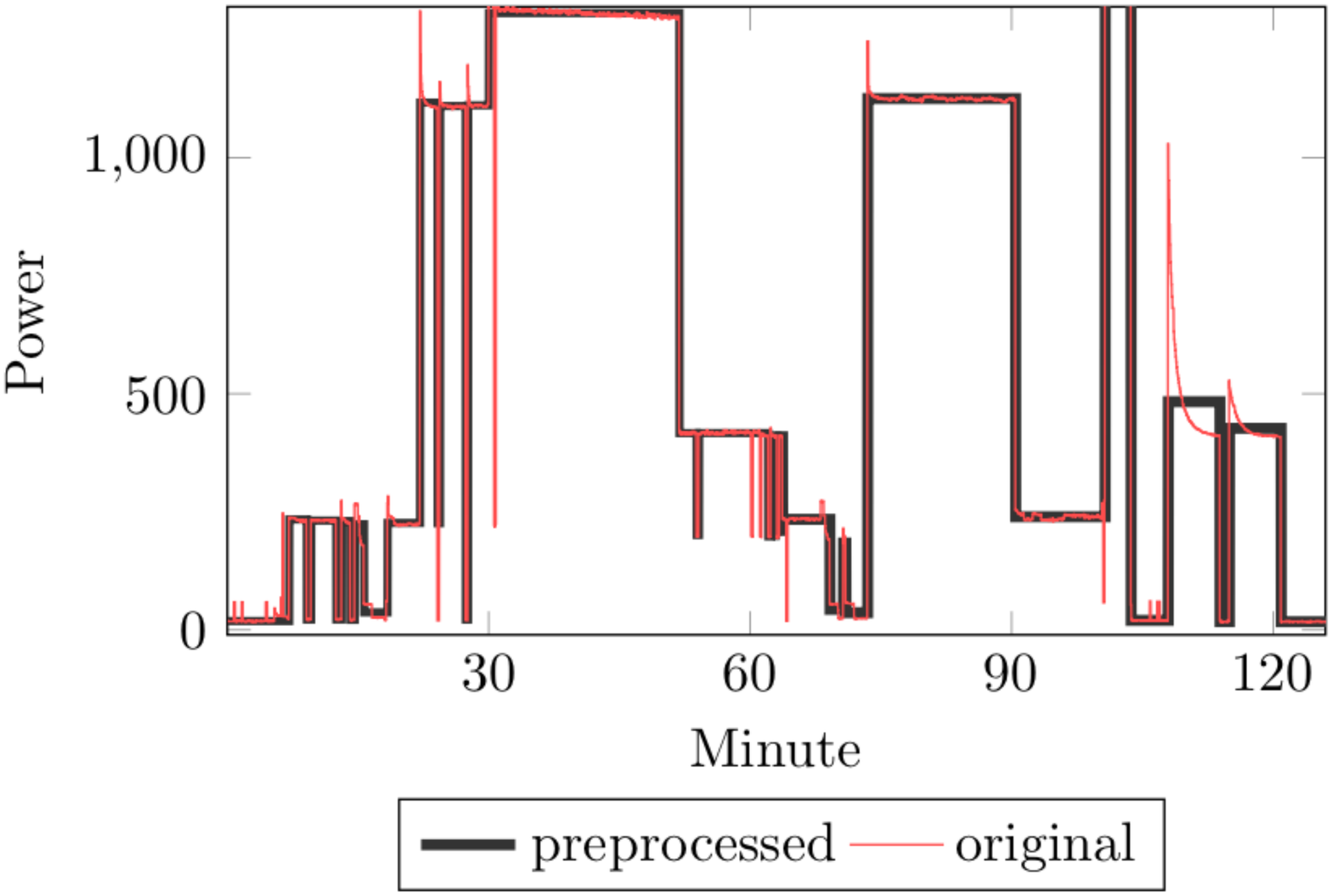}
\caption{Measured power signal vs. pre-processed power signal. The de-noising and smoothing filter remove high frequency fluctuations}
\label{fig:smoothedSignal}
\end{figure}

The presented approch has no {\em a priori} information about the number of appliances either the number of power states.
Thus, one big task is to detect power states and map this power states to states empirically identified as reference power states.
To evaluate the performance we took for each day the reference power state and the estimated power states.
Firstly, the power draw is de-noised and smoothed as presented in Figure \ref{fig:smoothedSignal}.
High frequency fluctuations are removed and a steady state of the power draw is reached by sharpening the edges.
This stage is helpful and necessary for the edge detection. 

%\begin{figure}[htpb]
%\centering
%		
%		\setlength\figureheight{4cm}
%		\setlength\figurewidth{7cm}
%		\input{zoomed.tikz}
%		\caption{Measured power signal vs. pre-processed power signal. The de-noising and smoothing filter remove high frequency fluctuations}
%		\label{fig:smoothedSignal}
%\end{figure}
After removing these power fluctuations, the power states are detected by edge detection and state clustering.
We tried to map the detected power states to the real ones by a simple distance measure.
As reference value we took a difference of $75W$.
The result are two numbers representing the number of appliance states able to mapped to reference power states and the number of power states which can not be mapped to reference power states.
The reference power states are empirically identified and therefore, we claim that appliance power states in group of not assignable power states are not necessarily false detected, but means that they belong to appliance states that are rarely occurring.
In total, the $9$ following appliance states are possible: $100, 200, 390, 800, 1100, 1500, 1650, 2600$ and $2720$ Watts.
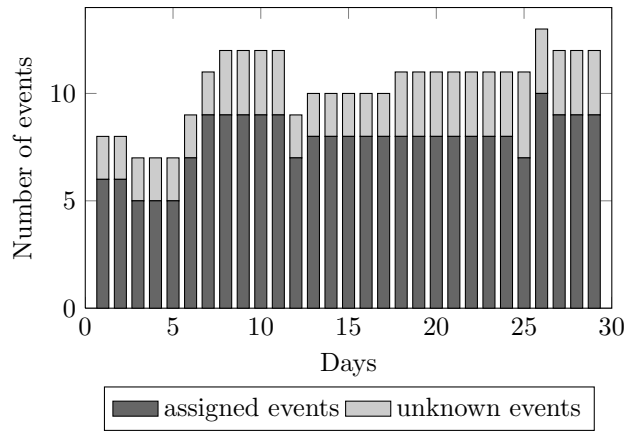
\begin{figure}[htpb]
\centering
\newlength\figureheight
		\newlength\figurewidth
		\setlength\figureheight{4cm}
		\setlength\figurewidth{7cm}
		% This file was created by matlab2tikz v0.4.7 running on MATLAB 8.1.
% Copyright (c) 2008--2014, Nico Schlömer <nico.schloemer@gmail.com>
% All rights reserved.
% Minimal pgfplots version: 1.3
% 
% The latest updates can be retrieved from
%   http://www.mathworks.com/matlabcentral/fileexchange/22022-matlab2tikz
% where you can also make suggestions and rate matlab2tikz.
% 
\begin{tikzpicture}

\begin{axis}[%
width=\figurewidth,
height=\figureheight,
ybar stacked,
bar width=0.0228571428571429\figurewidth,
area legend,
scale only axis,
xmin=0,
xmax=30,
xlabel={Days},
ylabel={Number of events},
ymin=0,
ymax=14,
legend style={at={(0.035408560311284,-0.4082359789138421)},anchor=south west,legend columns=3,draw=black,fill=white,legend cell align=left}
]
\addplot[draw=black,fill=white!40!black] plot table[row sep=crcr] {%
1	6\\
2	6\\
3	5\\
4	5\\
5	5\\
6	7\\
7	9\\
8	9\\
9	9\\
10	9\\
11	9\\
12	7\\
13	8\\
14	8\\
15	8\\
16	8\\
17	8\\
18	8\\
19	8\\
20	8\\
21	8\\
22	8\\
23	8\\
24	8\\
25	7\\
26	10\\
27	9\\
28	9\\
29	9\\
};
\addlegendentry{assigned events};

\addplot[draw=black,fill=black!20] plot table[row sep=crcr] {%
1	2\\
2	2\\
3	2\\
4	2\\
5	2\\
6	2\\
7	2\\
8	3\\
9	3\\
10	3\\
11	3\\
12	2\\
13	2\\
14	2\\
15	2\\
16	2\\
17	2\\
18	3\\
19	3\\
20	3\\
21	3\\
22	3\\
23	3\\
24	3\\
25	4\\
26	3\\
27	3\\
28	3\\
29	3\\
};
\addlegendentry{unknown events};

\end{axis}
\end{tikzpicture}%
		\caption{Number of power events detected and assignable to real power events. The number of real power events is 9.}
		\label{fig:detectedStates}
\end{figure}

\begin{figure}[htpb]
\centering
		\setlength\figureheight{4cm}
		\setlength\figurewidth{7cm}
		% This file was created by matlab2tikz v0.4.7 running on MATLAB 8.1.
% Copyright (c) 2008--2014, Nico Schlömer <nico.schloemer@gmail.com>
% All rights reserved.
% Minimal pgfplots version: 1.3
% 
% The latest updates can be retrieved from
%   http://www.mathworks.com/matlabcentral/fileexchange/22022-matlab2tikz
% where you can also make suggestions and rate matlab2tikz.
% 
%
% defining custom colors
\definecolor{mycolor1}{rgb}{0.90588,0.90588,0.90588}%
\definecolor{mycolor2}{rgb}{0.31373,0.31373,0.31373}%

\begin{tikzpicture}

\begin{axis}[%
width=\figurewidth,
height=\figureheight,
area legend,
scale only axis,
xmin=0,
xmax=3000,
xtick={ 100,  200,  390,  800, 1100, 1500, 1650, 2600, 2720},
xlabel={Power in W},
ymin=0,
ymax=18,
ylabel={Count of power Clusters},
  xticklabel style={
        inner sep=1pt,
        anchor=north east,
        rotate=70}
]
\addplot[ybar,bar width=0.0509714285714286\figurewidth,bar shift=0.0085714285714286\figurewidth,draw=mycolor1,fill=mycolor1] plot table[row sep=crcr] {%
100	17.9\\
200	17.9\\
390	17.9\\
800	17.9\\
1100	17.9\\
1500	17.9\\
1650	17.9\\
2600	17.9\\
2720	17.9\\
};
\addplot [color=black,solid,forget plot]
  table[row sep=crcr]{%
0	0\\
2500	0\\
};
\addplot[ybar,bar width=0.00195428571428571\figurewidth,bar shift=0.00122142857142857\figurewidth,draw=mycolor2,fill=mycolor2] plot table[row sep=crcr] {%
108.224724941877	7\\
134.825481475125	0\\
161.426238008372	0\\
188.02699454162	15\\
214.627751074867	14\\
241.228507608115	0\\
267.829264141362	0\\
294.430020674609	0\\
321.030777207857	0\\
347.631533741104	0\\
374.232290274352	0\\
400.833046807599	0\\
427.433803340847	11\\
454.034559874094	4\\
480.635316407342	0\\
507.236072940589	0\\
533.836829473837	0\\
560.437586007084	0\\
587.038342540332	0\\
613.639099073579	0\\
640.239855606826	2\\
666.840612140074	0\\
693.441368673321	0\\
720.042125206569	0\\
746.642881739816	0\\
773.243638273064	1\\
799.844394806311	2\\
826.445151339559	0\\
853.045907872806	3\\
879.646664406054	5\\
906.247420939301	2\\
932.848177472549	0\\
959.448934005796	0\\
986.049690539043	0\\
1012.65044707229	0\\
1039.25120360554	0\\
1065.85196013879	17\\
1092.45271667203	6\\
1119.05347320528	1\\
1145.65422973853	0\\
1172.25498627178	0\\
1198.85574280502	0\\
1225.45649933827	0\\
1252.05725587152	0\\
1278.65801240477	1\\
1305.25876893801	1\\
1331.85952547126	2\\
1358.46028200451	2\\
1385.06103853776	1\\
1411.661795071	0\\
1438.26255160425	2\\
1464.8633081375	1\\
1491.46406467075	0\\
1518.06482120399	10\\
1544.66557773724	15\\
1571.26633427049	1\\
1597.86709080373	2\\
1624.46784733698	1\\
1651.06860387023	3\\
1677.66936040348	1\\
1704.27011693672	0\\
1730.87087346997	0\\
1757.47163000322	0\\
1784.07238653647	0\\
1810.67314306971	0\\
1837.27389960296	0\\
1863.87465613621	0\\
1890.47541266946	0\\
1917.0761692027	0\\
1943.67692573595	0\\
1970.2776822692	0\\
1996.87843880245	0\\
2023.47919533569	0\\
2050.07995186894	0\\
2076.68070840219	0\\
2103.28146493544	0\\
2129.88222146868	0\\
2156.48297800193	0\\
2183.08373453518	0\\
2209.68449106843	0\\
2236.28524760167	0\\
2262.88600413492	0\\
2289.48676066817	0\\
2316.08751720142	0\\
2342.68827373466	0\\
2369.28903026791	0\\
2395.88978680116	0\\
2422.49054333441	0\\
2449.09129986765	1\\
2475.6920564009	0\\
2502.29281293415	0\\
2528.8935694674	0\\
2555.49432600064	2\\
2582.09508253389	0\\
2608.69583906714	0\\
2635.29659560039	0\\
2661.89735213363	4\\
2688.49810866688	2\\
2715.09886520013	4\\
2741.69962173338	1\\
};
\end{axis}
\end{tikzpicture}%
		\caption{Histogram of all detected power events over the observation window. The light-gray areas indicate the real power states identified empirically}
		\label{fig:detectedVsKnwon}
\end{figure}
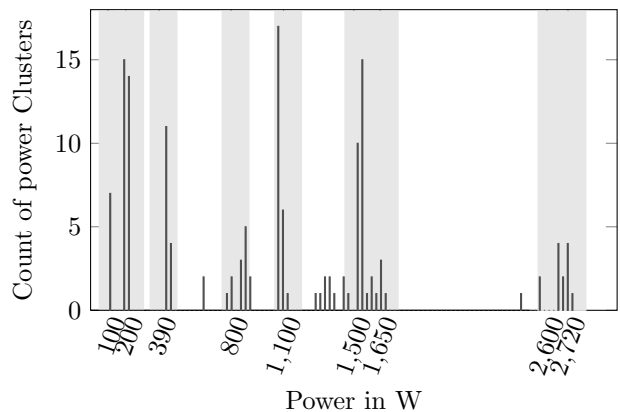
Figure \ref{fig:detectedStates} presents the results of detected and assignable/not assignable power states per day for $30$ consecutive days of $6$ different household appliances.
The graph shows that the number of detected power states is getting better and stable over days in operation.
The number of not assignable appliances stagnate since power states which are occurring frequently are not eliminated by the database update process.
Figure \ref{fig:detectedVsKnwon} illustrates the known appliance power states (gray area) against the detected power states (black bars) for a time duration of $30$ days.
Moreover, the power region between $1200W$ and $1500W$ indicates that the algorithm detects appliance state which were not identified empirically by hand.
\newline

\subsubsection{Load Disaggregation error on appliance level and for the total power draw}
The previous case study was evaluating how good the unsupervised state detection and appliance modelling process is working.
In the second case study the generated appliance models are used by the presented load disaggregation classification approach.
As described the appliance are modelled as \ac{HMM}s where the transition matrix is set a-priori.
To make the ground-truth appliance data comparable with the results of the load disaggregator we treat appliances with sufficiently similar power demands as the same virtual appliances.
Therefore, we empirically deployed three virtual appliances as representative for the evaluations of the used appliance set.
In the case of the load disaggregator we tried to assign the detected appliance states to the virtual appliances.
Power states not assignable are marked as \textit{unKnown}.

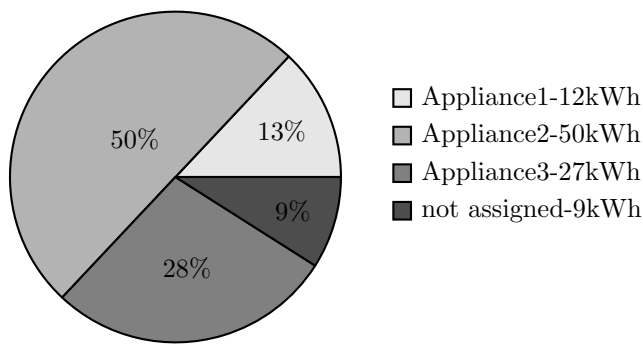
\begin{figure}
\centering
\begin{tikzpicture}
\pie[ radius =2.2,text = legend, color ={ black!10 , black!30 , black!50 , black!70}]{13/Appliance1-12kWh, 50/Appliance2-50kWh, 28/Appliance3-27kWh, 9/not assigned-9kWh}
\end{tikzpicture}
\caption{Pie chart of the estimated energy for the virtual appliances and power states which can be assigned to power states}
\label{fig:estimatedEnergy}

\end{figure}
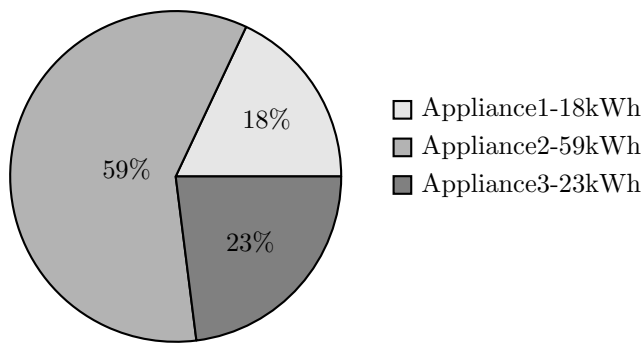
\begin{figure}
\centering
\begin{tikzpicture}
\pie[radius =2.2,text = legend, color ={ black!10 , black!30 , black!50 }]{18/Appliance1-18kWh, 59/Appliance2-59kWh, 23/Appliance3-23kWh}
\end{tikzpicture}
\caption{Pie chart of the ground truth energy for virtual appliance power states}
\label{fig:groundTruthEnergy}
\end{figure}

In Figure \ref{fig:estimatedEnergy} the power shares of the estimated energy per virtual appliance and not assignable power share are presented.
As comparison the ground truth pie chart of the power shares is illustrated in Figure \ref{fig:groundTruthEnergy}.

The results show that the estimation error is satisfying in which the error between the estimated ($100.728kWh$) and the real consumed energy ($101.51kWh$) is around $1\%$.
It is visible that most of the estimated appliance models and their power estimates can be assigned to the virtual appliances.
Around $9\%$ of the estimated energy can not be assigned to an virtual appliance.
As reason we claim the process how we assigned power states to the virtual appliances in which we chose an distance error of $75W$ from the detected power state and the real power state.

\section*{Acknowledgement}
This work was performed in the research cluster Lakeside Labs funded by the European Regional Development Fund, the Carinthian Economic Promotion Fund (KWF), and the state of Austria under grants 20214/22935/34445 (Smart Microgrid Lab) and 20214-23743-35469/35470 (Monergy).
We would like to thank Andrea Monacchi for useful discussion.

\section{Limitations \& Future Work} \label{sec:limitations}
The presented approach has several limitations which has to be improved by future research.
For example, the approach is considering two-state appliance models in which the state detection stage is already detecting all possible power states.
These power states can be part of an multi-state device.
Thus, we aiming to define rules and algorithms how to combine power events to multi-state appliance models.	
Further, we want also to improve the state detection process concerning long lasting power transients of appliances.
Some appliances and appliance types have transients which last for several seconds.
	This should be improved by advanced detection algorithms.
Finally, the problem of automatic appliance labelling to the correct appliance type is not considered yet.
Future work has to consider how to label detected appliances according to the detection history and general appliance type information as general operation duration or occurrence frequency per day.
\section{Conclusion}\label{sec:conclusion}
In this paper an unsupervised approach to solve the problem to disaggregate appliance power draws from the aggregated power load was presented.
The approach autonomously detects the power states of the used appliances. It improves the saved appliance models in operation and updates the appliance database by adding new appliance models and by removing rarely occurring appliance models.
The detected appliance models can be used by the load disaggregator to estimate the appliance states. The estimation results are promising in particular because of the low amount on not assignable energies and the good overall estimation result.

The models for each appliance are learned in run-time of the algorithm.
The algorithm contains a preprocessing stage to de-noise and smooth the aggregated power draw in a way to be able to detect sharp and significant power edges.
Only with the knowledge of the power edges appliance models as on/off appliances are established used by the load disaggregator based on particle filtering.
The approach is evaluated on real measurement data where our results emphasizes the proposed \ac{NILM} approach as a very promising approach.

The number of detected appliance states and the corresponding disaggregation result is sufficient and satisfying and had been achieved without appliance information from the user. Future work will aim at multiple appliance modelling and automatic appliance labelling.

\bibliographystyle{alpha}
\bibliography{dominik}

\newcommand{\etalchar}[1]{$^{#1}$}
\begin{thebibliography}{PGWR14}

\bibitem[BKC{\etalchar{+}}14]{Beckel2014}
Christian Beckel, Wilhelm Kleiminger, Romano Cicchetti, Thorsten Staake, and
  Silvia Santini.
\newblock The eco data set and the performance of non-intrusive load monitoring
  algorithms.
\newblock In {\em Proceedings of the 1st ACM Conference on Embedded Systems for
  Energy-Efficient Buildings}, BuildSys '14, pages 80--89, New York, NY, USA,
  2014. ACM.

\bibitem[BMI{\etalchar{+}}12]{barker2012smart}
Sean Barker, Aditya Mishra, David Irwin, Emmanuel Cecchet, Prashant Shenoy, and
  Jeannie Albrecht.
\newblock Smart*: An open data set and tools for enabling research in
  sustainable homes.
\newblock 2012.

\bibitem[CMB13]{Carlson2013132}
Derrick~R. Carlson, H.~Scott Matthews, and Mario Berges.
\newblock One size does not fit all: Averaged data on household electricity is
  inadequate for residential energy policy and decisions.
\newblock {\em Energy and Buildings}, 64(0):132 -- 144, 2013.

\bibitem[EBE14]{Egarter2014}
Dominik Egarter, Venkata~Pathuri Bhuvana, and Wilfried Elmenreich.
\newblock {PALDi}: Online load disaggregation via particle filtering.
\newblock {\em {IEEE} Transactions on Instrumentation and Measurement}, 2014.

\bibitem[EE12]{elmenreich:wises12}
W.~Elmenreich and D.~Egarter.
\newblock Design guidelines for smart appliances.
\newblock In {\em Proceedings of the 10th International Workshop on Intelligent
  Solutions in Embedded Systems}, 2012.

\bibitem[EPE14]{prokop:14}
D.~Egarter, C.~Prokop, and W.~Elmenreich.
\newblock Load hiding of household's power demand.
\newblock In {\em Proc. IEEE International Conference on Smart Grid
  Communications (SmartGridComm'14)}, Venice, Italy, 2014.

\bibitem[Har92]{Hart1992}
G.W. Hart.
\newblock {Nonintrusive appliance load monitoring}.
\newblock {\em Proceedings of the IEEE}, 80(12):1870--1891, 1992.

\bibitem[JW13]{Johnson2013}
Matthew~J. Johnson and Alan~S. Willsky.
\newblock Bayesian nonparametric hidden semi-markov models.
\newblock {\em J. Mach. Learn. Res.}, 14(1):673--701, February 2013.

\bibitem[KJ11]{kolter-kdd-2011}
J.~Zico Kolter and Matthew~J. Johnson.
\newblock {REDD: A Public Data Set for Energy Disaggregation Research}.
\newblock In {\em Proceeding of the SustKDD Workshop on Data Mining
  Applications in Sustainability}, 2011.

\bibitem[KJ12]{Zico2012}
Zico Kolter and Tommi Jaakkola.
\newblock Approximate inference in additive factorial {HMMs} with application
  to energy disaggregation.
\newblock In {\em Proceedings of the International Conference on Artifical
  Intelligence and Statistics}, 2012.

\bibitem[KMA{\etalchar{+}}11]{Kim2011}
H~Kim, M~Marwah, M~F Arlitt, G~Lyon, and J~Han.
\newblock {Unsupervised Disaggregation of Low Frequency Power Measurements}.
\newblock In {\em Proceedings of the 11th {SIAM} International Conference on
  Data Mining}, 2011.

\bibitem[MEDT13]{monacchi2013Nov}
Andrea Monacchi, Wilfried Elmenreich, Salvatore D'Alessandro, and Andrea~M.
  Tonello.
\newblock Strategies for energy conservation in {C}arinthia and
  {F}riuli-{V}enezia {G}iulia.
\newblock In {\em Proc. of the 39th Annual Conference of the IEEE Industrial
  Electronics Society}, 2013.

\bibitem[MEE{\etalchar{+}}14]{Andrea2014}
Andrea Monacchi, Dominik Egarter, Wilfried Elmenreich, Salvatore
  D’Alessandro, and Andrea~M. Tonello.
\newblock {GREEND}: an energy consumption dataset of households in {I}taly and
  {A}ustria.
\newblock In {\em Proceedings of IEEE International Conference on Smart Grid
  Communications ({SmartGridComm})}, 2014.

\bibitem[MPB{\etalchar{+}}13]{makonin2013ampds}
Stephen Makonin, Fred Popowich, Lyn Bartram, Bob Gill, and Ivan~V. Bajic.
\newblock {AMPds: A Public Dataset for Load Disaggregation and Eco-Feedback
  Research}.
\newblock In {\em Electrical Power and Energy Conference (EPEC), 2013 IEEE},
  pages 1--6, 2013.

\bibitem[Pat12]{pattem2012}
S.~Pattem.
\newblock Unsupervised disaggregation for non-intrusive load monitoring.
\newblock In {\em Machine Learning and Applications (ICMLA), 2012 11th
  International Conference on}, volume~2, pages 515--520, Dec 2012.

\bibitem[PGWR14]{Parson2014}
Oliver Parson, Siddhartha Ghosh, Mark Weal, and Alex Rogers.
\newblock An unsupervised training method for non-intrusive appliance load
  monitoring.
\newblock {\em Artificial Intelligence}, 217(0):1 -- 19, 2014.

\bibitem[SMR13]{Shao2013}
Huijuan Shao, Manish Marwah, and Naren Ramakrishnan.
\newblock A temporal motif mining approach to unsupervised energy
  disaggregation: Applications to residential and commercial buildings.
\newblock In {\em Proceedings of the Twenty-Seventh {AAAI} Conference on
  Artificial Intelligence, July 14-18, 2013, Bellevue, Washington, {USA.}},
  2013.

\bibitem[ZBZ11]{Zia2011}
T.~Zia, D.~Bruckner, and A.~Zaidi.
\newblock A hidden markov model based procedure for identifying household
  electric loads.
\newblock In {\em Proceedings of Annual Conference on {IEEE} Industrial
  Electronics Society ({IECON})}, 2011.

\bibitem[ZGIR12]{Zoha2012}
Ahmed Zoha, Alexander Gluhak, Muhammad~Ali Imran, and Sutharshan Rajasegarar.
\newblock Non-intrusive load monitoring approaches for disaggregated energy
  sensing: A survey.
\newblock {\em Sensors}, 12(12):16838--16866, 2012.

\bibitem[ZGNI13]{Zoha2013}
A.~Zoha, A.~Gluhak, M.~Nati, and M.A. Imran.
\newblock Low-power appliance monitoring using factorial hidden markov models.
\newblock In {\em Proceedings of {IEEE} Eighth International Conference on
  Intelligent Sensors, Sensor Networks and Information Processing}, 2013.

\bibitem[ZR11]{Zeifman2011}
M.~Zeifman and K.~Roth.
\newblock Nonintrusive appliance load monitoring: Review and outlook.
\newblock {\em {IEEE} Trans. Consum. Electron.}, 57(1):76 --84, february 2011.

\end{thebibliography}

\end{document}